\documentclass[preprint]{aastex}
\usepackage{CJK}
\usepackage{amsmath}
\usepackage{graphicx}
\usepackage{graphics}
\usepackage[dvips]{color}
\newcommand{\kms}{{km~s$^{-1}\ $}}
\def\J1523{\mbox{SDSS~J1523+3914}}

\begin{document}
\title{Discovery of Extremely Broad Balmer Absorption Lines in SDSS J152350.42+391405.2}

\author{ Shaohua~Zhang\altaffilmark{1}, Hongyan~Zhou\altaffilmark{1,2}, Xiheng Shi\altaffilmark{1,3}, 
Xinwen Shu\altaffilmark{4}, Wenjuan Liu\altaffilmark{1,2}, Tuo~Ji\altaffilmark{1}, Peng~Jiang\altaffilmark{1,5},
Luming Sun\altaffilmark{1,2}, Junyan Zhou\altaffilmark{1,2} and Xiang Pan\altaffilmark{1,2}}

\altaffiltext{1}{Polar Research Institute of China, 451 Jinqiao Road, Shanghai, 200136, China;
zhangshaohua@pric.org.cn; zhouhongyan@pric.org.cn}
\altaffiltext{2}{Key Laboratory for Researches in Galaxies and Cosmology, Department of Astronomy,
University of Sciences and Technology of China, Chinese Academy of Sciences, Hefei, Anhui, 230026, China}
\altaffiltext{3}{National Astronomical Observatories, Chinese Academy of Sciences, Beijing, 100012, China}
\altaffiltext{4}{Department of Physics, Anhui Normal University, Wuhu, Anhui, 241000, China}
\altaffiltext{5}{School of Astronomy and Space Science, Nanjing University, Nanjing 210093, China}

\begin{abstract}
We present the discovery of Balmer line absorption from H$\alpha$ to H$\gamma$ in an iron low-ionization broad absorption line (FeLoBAL) quasar SDSS J152350.42+391405.2 (hereafter \J1523), by the quasi-simultaneous optical and near-infrared spectroscopy.
The Balmer line absorption is at $z_{absor} = 0.6039\pm0.0021$ and blueshifted by $v=10,353$ \kms with respect to the Balmer emission lines. All Balmer BALs have uniform absorption profile with the widths of $\Delta v \sim$12,000 \kms. We also found the absorption trough in \ion{He}{1}*$\lambda$10830 with the same velocity and width in the $H$-band TripleSpec spectrum of \J1523. This object is only the tenth active galactic nucleus known to exhibit non-stellar Balmer absorption, and also the case with the highest velocity and broadest Balmer absorption lines which have ever been found.  A CLOUDY analysis shows that the absorbers require an gas density of ${\rm log_{10}}~n_{\rm e}~({\rm cm^{-3}})=9$ and an ionization parameter of ${\rm log_{10}}~U=-1.0$. They locate at a distance of $\sim0.2$ pc from the central ionizing source which is slightly farther than that of BELRs. Furthermore, \J1523 is one of the brightest Balmer-BAL quasar ever reported, with unique iron absorption variations, making it as the most promising candidate for follow up high-resolution spectroscopy, multi-band observations, and long-term monitoring.

\end{abstract}


\section{Introduction}
Evidences accumulated in the past decade point to the phenomenon that feedback from active galactic
nuclei (AGNs) plays a crucial role in galaxy formation and evolution.
Outflows in AGNs connect the central supermassive black holes (SMBHs) to their host galaxies, and regulate their co-evolution
(Granato et al. 2004; Scannapieco \& Oh 2004; Hopkins et al. 2008; also see Antonuccio-Delogu \& Silk 2010 for a recent review). 
Observationally, the most direct and obvious performance of AGN outflows is the high-speed blueshifted broad absorption lines (BALs). 
Because of the high fraction in optically selected quasars (10\%-20\%), 
high-ionization BAL (HiBAL) quasars are widely studied, with strong absorption troughs in 
high ionization ions such as \ion{N}{5}, \ion{C}{4}, \ion{Si}{4}, and \ion{O}{6}, up to
a velocity of $v\sim 0.2 \rm c$ (e.g., Weymann et al. 1991; Trump et al. 2006; Gibson et al. 2009). 
BALs are also detected occasionally ($\sim 15\%$ in BAL quasars and $\sim 2\%$ in all quasars) 
in low ionization species such as \ion{Al}{3} and \ion{Mg}{2}, 
named as low-ionization BALs (LoBALs) (e.g., Weymann et al. 1991; Reichard et al. 2003; Zhang et al. 2010).
Moreover, FeLoBAL quasars have been observed to exhibit the broad absorption troughs in \ion{Fe}{2} and/or 
\ion{Fe}{3}, which are rare and only found in only $\sim 15\%$ of LoBAL quasars (Hall et al. 2002). 

Statistical studies are widely used to explain the BAL phenomenon. 
BAL quasars on average have red continua (Weymann et al. 1991; Richards et al. 2003; Zhang et al. 2010, 2014) 
 weak X-ray emission (e.g., Green et al. 1995; Brinkmann et al. 1999; Wang et al. 1999; Brandt et al. 2000; 
Gallagher et al. 2002, 2006; Fan et al. 2009), are more frequently detected in quasars with higher Eddington ratio 
and higher luminosity (Ganguly et al. 2007; Zhang et al. 2010, 2014).
Furthermore, the outflow velocity and strength are tied to the properties of quasars.
Observed outflow velocities increase with the blueness of UV spectral slope, 
the enhancement of black hole accretion and the equivalent width of \ion{He}{2} emission
(Hamann 1998; Laor \& Brandt 2002; Gungly et al. 2007; Misawa et al. 2007; Baskin et al. 2013; Zhang et al. 2014).
In particular, the minimum velocity of absorption is even more strongly correlated 
with UV spectral slope than the maximum velocity (Zhang et al. 2014).
These findings indicate the primary role of radiatively driven winds in the outflow phenomenon, 
and the importance of the spectral energy distribution shape (SED) in governing the dynamics of outflows.
Meanwhile, all outflow parameters dramatically and monotonically increase with hot dust emission.
These correlations can be more naturally interpreted as the dusty outflow scenario 
rather than the dust-free outflow scenario, where the dust is intrinsic to the outflows or interaction with torus clouds
(Wang et al. 2013; Zhang et al. 2014).

On the other hand, the study of variation of BAL troughs and the associated rare absorption systems, i.e., LoBALs, and unusual BALs 
(mostly FeLoBALs) can provide a new and more effective perspective to understand the physical conditions, 
locations and origins of the absorbers, and constraint on the outflow mechanism, which gradually becomes a hot topic of the BAL research 
(e.g., Hall et al. 2002, 2011; Zhou et al. 2006; Lundgren et al. 2007; Gibson et al. 2008, 2010; Krongold et al. 2010; 
Zhang et al. 2011, 2015a, 2015b; Capellupo et al. 2012; Vivek et al. 2012, 2014; Filiz Ak et al. 2013; Welling et al. 2014).
Currently, one of the rarest known BALs is non-stellar Balmer line absorption.
It has previously been reported only in nine objects as follows: NGC 4151\footnote{The \ion{Fe}{2} absorption spectrum 
of NGC 4151 was taken in 1999 July and published in Kraemer et al. (2001).} (Hutchings et al. 2002), 
SDSS J112526.12+002901.3 (Hall et al. 2002; Shi et al. in pre.), 
SDSS J083942.11+380526.3 (Aoki et al. 2006), SDSS J125942.80+121312.6 (Hall 2007; Shi et al. in pre.), 
SDSS J102839.11+450009.4 (Wang et al. 2008), SDSS J172341.10+555340.5 (Aoki 2010), LBQS 1206+1052 (Ji et al. 2012), 
SDSS J222024.59+010931.2 (Ji et al. 2013) and SDSS J112611.63+425246.4 (Wang \& Xu 2015).
However, strictly speaking, more than half of them cannot be classified as Balmer BALs based on the criteria for BALs 
(see Table 1 of Zhang et al. (2010) for a summary and comparison), as their absorption widths are narrower than 1000 \kms.

In this paper, we report a quasar (SDSS J152350.42+391405.2, hereafter \J1523) 
 with a emission redshift of $z_{\rm emi}=0.6612\pm0.0022$. This object shows 
the broadest and highest blueshifted velocity Balmer BALs known so far in the quasars,
suggesting the strong, high speed and high column density outflow materials in the nuclear region. 
The organization of this paper is as follows. The data we used will be described in Section 2. 
We will fit the spectrum and analyze the Balmer BALs in Section 3, and discuss the properties 
and possible origins of BALs in Section 4. A summary of our results will be given in Section 5.
Throughout this paper, we adopt the cold dark matter `concordance' cosmology
with H0 = 70 km s$^{-1}$Mpc$^{-1}$, $\Omega_{\rm m} = 0.3$, and $\Omega_{\Lambda} = 0.7$.

\section{OBSERVATIONS}%

\begin{deluxetable}{cccc}
\tabletypesize{\scriptsize}
\tablecaption{Photometric Data of \J1523
\label{tab1} }
\tablewidth{0pt}
\tablehead{
\colhead{Band}&
\colhead{FLUX/Magnitude}&
\colhead{Date-Observation} &
\colhead{Survey/Telescope}
}
\startdata
20cm &$~~4.07\pm0.13$ mJy  &1994/08/13 & FIRST\\
20cm &$~~4.25\pm0.26$ mJy  &1998/04/16 & NVSS\\
$u$  &$18.234\pm0.026$ mag &2003/02/11 & SDSS\\
$g$  &$16.938\pm0.032$ mag &2003/02/11 & SDSS\\
$r$  &$16.680\pm0.023$ mag &2003/02/11 & SDSS\\
$i$  &$16.468\pm0.031$ mag &2003/02/11 & SDSS\\
$z$  &$16.339\pm0.023$ mag &2003/02/11 & SDSS\\
$J$  &$15.362\pm0.053$ mag &1999/05/24 & 2MASS\\
$H$  &$14.875\pm0.066$ mag &1999/05/24 & 2MASS\\
$K_s$&$13.866\pm0.059$ mag &1999/05/24 & 2MASS\\
$W1$ &$12.011\pm0.027$ mag &2010/01/19, 07/17, 07/22 & WISE\\
$W2$ &$10.819\pm0.026$ mag &2010/01/19, 07/17, 07/22 & WISE\\
$W3$ &$~8.431\pm0.033$ mag &2010/01/19, 07/16, 07/22 & WISE\\
$W4$ &$~6.505\pm0.038$ mag &2010/01/19, 07/16, 07/22 & WISE\\
$V$  &     --              &2005/04/20--2013/09/28  & Catalina   \\
$V$  &     --              &2015/04/20,05/13 & BSST
\enddata
\end{deluxetable}

\begin{figure*}[tbp]
\epsscale{0.6} \plotone{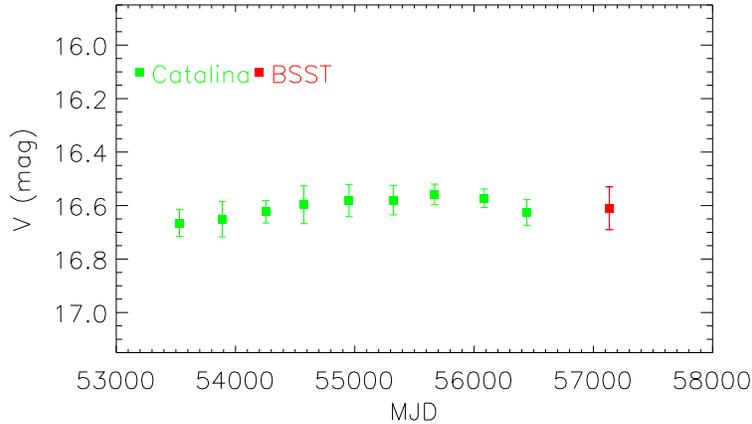}
\caption{Light curve of \J1523 in the $V$-band.
Variability amplitudes are calculated using the formalism in Ai et al. (2010).
The long-term variance of observed magnitudes $\Sigma_{\rm V}$ is $0.06\ {\rm mag}$ and
the intrinsic amplitude $\sigma_{\rm V}$ ($=\sqrt{\Sigma_{\rm V}^2-\xi^2}$) equals to zero.
Here, $\xi^2$ is measurement errors, which is estimated from the errors of observed individual magnitudes $\xi_i$, as 
$\xi^2=\frac{1}{N} \sum\limits_{i=1}^{N} \xi_i^{2}$.}
\label{f1}
\end{figure*}

\J1523 is very bright with a Galactic extinction corrected magnitude of 16.59 at $r$-band.
The FIRST (the Faint Images of the Radio Sky at Twenty cm, Becker et al. 1995) 
and NVSS (NRAO VLA Sky Survey; Condon et al. 1998) surveys 
show that there is no radio variation at 1.4 GHz,
with the peak flux of $4.06\pm0.13$ and $4.25\pm0.26$ mJy respectively.
At the high energy band, its X-ray emission is very weak and not detected by 
XMM-Newton and Chandra X-ray Observatories. 
The SED from ultraviolet  (UV) to middle-infrared (MIR), comes from 
the data taken with the Sloan Digital Sky Survey (SDSS, York et al. 2000),
the two micron all sky survey (2MASS; Skrutskie et al. 2006) and the Wide-field Infrared Survey 
Explorer (WISE; Wright et al. 2010). The UV emission is attenuated by the strong \ion{Fe}{2} and \ion{Mg}{2} absorption. 
Meanwhile, this object is also monitored by the Catalina Surveys\footnote{The Catalina Web site 
is http://nesssi.cacr.caltech.edu/DataRelease/.
The observed magnitudes of the Catalina Surveys are combined in one observing season,
and translated to the V-band magnitude of Johnson-Cousins filter system following the method described in Landolt (2009).}
(Drake et al. 2014) for the nine years (278 epochs), beginning in April 2005. 
Figure \ref{f1} presents the $V$-band light curve of \J1523 (green squares), which shows very weak long-term variability 
with large measurement errors.
The Antarctic bright star survey telescope (BSST; Tian et al., in pre.)\footnote{BSST is a 30 cm 
automatic optical photometric telescope which is designed for Chinese Antarctic Kunlun station,
and it will be deployed in Antarctic to discover and explore extrasolar planets at Dome A, Antarctica.}
also gives 8 epoch photometric observations in the $V$-band  
at the Lijiang observational station of Yunnan observatories.
The latest monitoring result (red square) confirms there is no apparent change in the optical continuum for \J1523.
The photometric data are summarized in Table \ref{tab1}. 
\figurenum{1}

\figurenum{2}
\begin{figure*}[tbp]
\epsscale{1.} \plotone{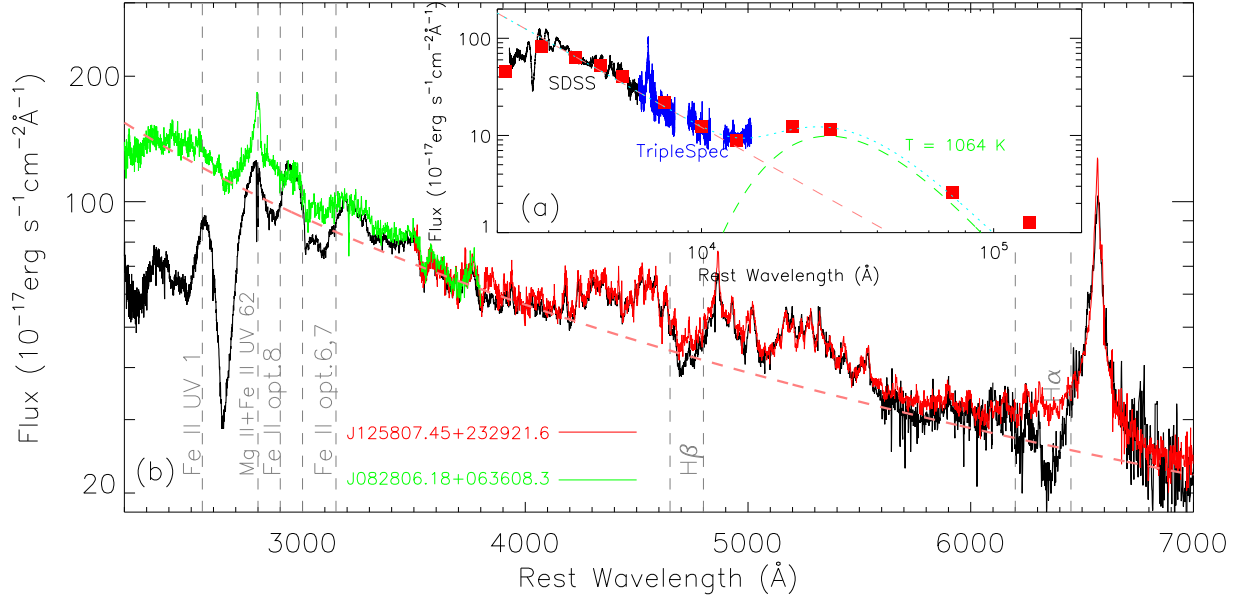}
\caption{Panel (a): We show the broad band spectral energy distribution (SED) of \J1523~from ultraviolet
to middle-infrared by red squares, the spectra of SDSS and TripleSpec by black and blue curves, 
the power-law continuum by pink and the blackbody radiation by green.
Panle (b): We show the spectral details of \J1523 from 2300 \AA~to 7000 \AA~by black curves, 
its FBQS spectrum by blue curves, and two quasars from the SDSS by green and red curves, 
whose spectra follow the continuum slop and the emission peaks of \ion{Fe}{2} multiplets of \J1523,
except BAL troughs.} 
\label{f2}
\end{figure*}

The optical spectrum of \J1523 was first taken with the 3.5 m telescope on Apache Point Observatory (APO) 
at May 26, 1997 in the FIRST Bright Quasar Survey (FBQS, White et al. 2000). 
The APO spectrum has a wavelength coverage from 3650 to 10000 \AA~at $\sim10~\rm \AA$ resolution.
\J1523 was also considered as a quasar candidate from the spectroscopy taken with the SDSS 2.5m telescope on May 6, 2003 and May 28, 2012.
The first observation was recorded into the SDSS Data Release 5 (DR5; Adelman-McCarthy et al. 2007). Two 5400s exposures were taken with
the original SDSS spectrographs, which provide a high signal-to-noise (SNR) spectrum at the resolution 
$R\sim1800$ and the wavelength coverage from 3800 to 9200 \AA~(Stoughton et al. 2002).
The second observation was taken in the Baryon Oscillation Spectroscopic
Survey (BOSS; Dawson et al. 2013) and pubilshed in the SDSS Tenth Data Release (DR10, Ahn et al. 2013),
The BOSS spectrum has a wider wavelength range covering 361 nm - 1014 nm
with a resolution of 1300 at the blue side and 2600 at the red side, respectively.
This instrument is fed by smaller optical fibers, each subtending 2'' on the sky (Smee et al. 2013).

The near-infrared (NIR) spectrum of \J1523 was taken with the TripleSpec spectrograph of Hale 200-inch telescope (P200)
at Palomar Observatory on April 16, 2012. 
\J1523\ was exposed four times, each for 240 seconds.  
TripleSpec (Wilson et al. 2004) is a cross-dispersed NIR spectrograph that provides simultaneous
wavelength coverage from 0.9 to 2.46 microns at a resolution of 1.4-2.9 \AA.
The raw data were processed using IDL-based Spextool software(Vacca et al. 2003 and Cushing et al. 2004).
There are two gaps in infrared spectrum around 1.35 microns and 1.85 microns due to the atmosphere transmissivity.
Fortunately, the redshifted H$\alpha$ emission line is detected with the TripleSpc at $J$-band.
Since the time interval between the SDSS DR10 and P200 TripleSpec observations is just 12 days in observed frame,
the data are considered as quasi-simultaneous which we will use in the following analysis.

After corrected for the Galactic reddening of $E(B-V)$ =0.021 mag (Schlegel et al. 1998),
we transformed the photometric data as well as optical and
NIR spectrum into the rest frame with its emission redshift, which are shown in
red squares, black and blue curves in panle (a) of Figure \ref{f2}, respectively.
The TripleSpec spectrum, after correcting for the aperture effect,
agrees very well with the 2MASS photometric data.
The SDSS and TripleSpec continuum spectra and WISE photometric data suggest two basic components. 
One is a singel power-law, $f_{\lambda}\propto\lambda^{-1.70}$, 
representing the nuclear spectrum from the accretion disk (the pink dashed line). 
 It is estimated from continuum windows ([3790, 3810]\AA, [5600, 5630]\AA, [6950, 7000]\AA)
that are not seriously contaminated by emission lines (e.g., Forster et al. 2001).
Another is the NIR emission bump at $W2$-band corresponding to the blackbody radiation 
from the hot dust with $T = 1064$ K (green dashed line).

\section{ANALYSIS}

\J1523 is a typical BAL quasar, and first classified as a radio-detected BAL quasar with $BI\sim 3700$ \kms~by Becker et al. (2000). 
It is also contained in the SDSS DR5 low-redshift \ion{Mg}{2} BAL quasar sample (Zhang et al. 2010).
The measured maximum and minimum velocities of \ion{Mg}{2} BAL trough are -20,000 and -11,500 \kms, 
and the absorption index $AI$ is 3227 \kms. We note that the absorption intensity of \ion{Mg}{2} BAL listed 
in the above literature is likely underestimated as the absorbing troughs of \ion{Mg}{2} and \ion{Fe}{2} UV62 overlap each other, 
and can not be discriminated easily. In Zhang et al. (2015b), \J1523 was reidentified to be an `overlapping-trough' 
FeLoBAL quasar because of almost no continuum windows below Mg II and causing overlapping \ion{Fe}{2} absorption troughs.

\subsection{Absorption of Balmer Lines }
In panel (b) of Figure \ref{f2}, we show the spectral details of \J1523 from 2300 \AA~ to 7000 \AA.
Several regions of the spectrum are absorbed and significantly below the nuclear continuum. 
Except BALs of \ion{Mg}{2} and \ion{Fe}{2}, Balmer BALs can be also found by visual examination.
We further selected two quasars (SDSS J082806.18+063608.3 and SDSS J125807.45+232921.6) using the pair-matching method (Zhang et al. 2014; Liu et al. 2015),
which can match the continuum slop and \ion{Fe}{2} emission multiplets of \J1523\ except for the absorbing troughs.
Compared with the spectra of the two quasars (green and red curves), 
BAL troughs of H$\alpha$ and H$\beta$ are evident. 
 We marked the potential BAL absorption regions by gray dashed lines.

In order to exactly measure the absorption troughs of Balmer lines, we simultaneously fit the continuum and emission lines
in the H$\beta$ and H$\alpha$ regions using the code of Dong et al. (2008).
In brief, the optical continuum from 4000 \AA~ to 7000 \AA~ is approximated by a single power-law ($F_{\lambda} \propto \lambda^{\beta}$).
The value of the slope estimated from continuum windows is took as initial values of $\beta$.
Indeed, we also try to use a broken power-law with a break wavelength of 5600 \AA,
i.e., $a_{1} \lambda ^{\beta_1}$ for the H$\beta$ region 
and $a_{2} \lambda ^{\beta_2}$ for the H$\alpha$ region. 
However, we get the almost same values of ${\beta_1}$ and ${\beta_2}$. 
That means  a single power-law is enough to model the continuum.
\ion{Fe}{2} multiplets, both broad and narrow, are modeled using the \mbox{I Zw 1} template provided by V{\'e}ron-Cetty et al. (2004).
Emission lines are modeled as multiple Gaussians: two Gaussians for broad Balmer lines, 
one Gaussian for [\ion{O}{3}]. There are no significant narrow emission lines 
in the spectra of \J1523, thus we do not add Gaussians for narrow emission lines.
Additionally, we assume that the [\ion{O}{3}]$\lambda\lambda$4959,5007 doublets 
have the same redshift and profile and their flux ratio is fixed to theoretical value.
We notice that the spectrum of SDSS J125807.45+232921.6 can cover the blue wing ($v\lesssim5000$ \kms) and red wing 
of broad H$\alpha$ line in \J1523 (Figure \ref{f2}, panel (b)), that means H$\alpha$ emission is likely unaffected by the BAL absorption.
Thus, we just mask the potential BAL absorption regions (gray dashed lines) in the fitting.

\begin{figure*}[tbp]
\epsscale{1.} \plotone{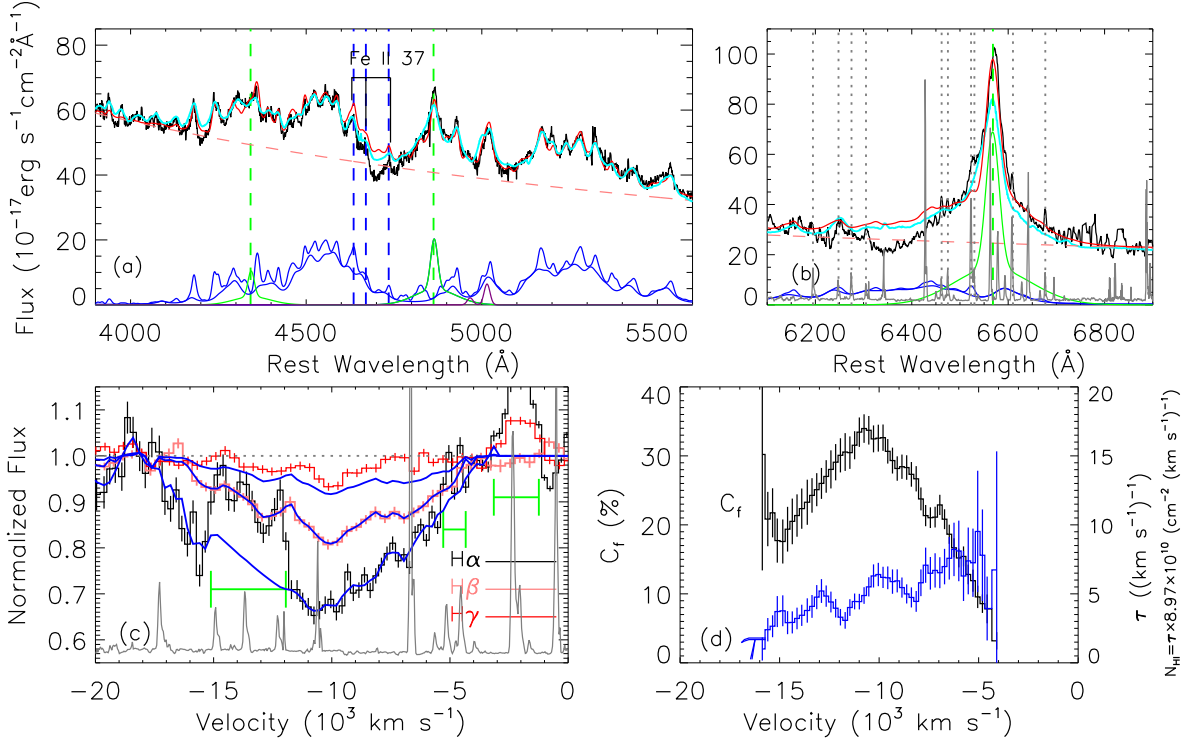}
\caption{Panel (a) and (b): Observed spectra of H$\beta$ and H$\alpha$ regimes 
with the best-fit models for \J1523. The models are described in detail in Section 2, 
In each panel, we plot the observed spectrum in black curves, power-law continuum in pink, 
broadened \ion{Fe}{2} template in blue, Gaussian Balmer emission lines in green, and the
model sum in red. The cyan curves show the best-fit model with \ion{Fe}{2} and Balmer emission
template obtained from IRAS 07598+6508. 
Panel (c): Normalized spectra obtained by dividing the observed spectra by the sum of the modeled continuum and 
emission lines in the velocity space. We also overplot the theoretical absorption troughs obtained 
from the absorption depths and the covering factors of \ion{H}{1} (Panel (d)) by blue curves.
The velocity regions of H$\alpha$ BAL polluted by Sky lines are marked by green lines.
The gray dotted line means the value equals 1.
Panel (d): Percentage covering factor and the optical depth (the column density of \ion{H}{1}) as a function of velocity in Balmer absorption troughs.
Sky line spectrum (not scaled) is plotted as grey line in panel (b) and (c) for comparison.}\label{f3}
\end{figure*}

In top panels of Figure \ref{f3}, we show the rest-frame spectra of the SDSS and TripleSpec in black curves and our best-fit models 
in red curves. A more accurate continuum around H$\alpha$ and H$\beta$,  
$f_{\lambda}\propto\lambda^{-1.6816}$, is shown in pink dashed line. The blue solid lines show the broad and narrow 
components of optical \ion{Fe}{2} emission and the green curves represent the three strong Balmer lines.
In panel (a), the strong narrow \ion{Fe}{2} 37 multiplets are present in H$\beta$ BAL trough.
This implies that the outflow winds may only obscure the nuclear continuum and broad emission lines, but not the narrow emission lines.
This speculation is consistent with our detailed calculation in \S 4.2.
Furthermore, it can be seen from the normalized spectra of  H$\alpha$, H$\beta$ and H$\gamma$ (Figure \ref{f3}, panel (c))
that the  H$\alpha$ trough is polluted by the sky lines. The latter is plotted as the grey line in panel (b) of Figure \ref{f3} for comparison.
That is more remarkable in the normalized spectrum of H$\alpha$ in velocity space, 
the corresponding polluted velocity regions are marked by green lines (panel (c)). 
The `true' normalized spectrum in these polluted regions is approximately given through interpolation of those in unpolluted regions. 
Absorption parameters of Balmer BAL troughs  are listed in Table \ref{tab2}.

\begin{deluxetable}{lccc}
\tabletypesize{\scriptsize}
\tablecaption{Absorption Line Parameters of \J1523
\label{tab2} }
\tablewidth{0pt}
\tablehead{
\colhead{} &
\colhead{H$\alpha$ } &
\colhead{H$\beta$  } &
\colhead{H$\gamma$ } 
}
\startdata
 Wavelength(\AA)			& 6564.41        & 4862.68     & 4341.68    \\
 $f_{\rm ij}$ $^{a}$		& 0.6400         & 0.1190      & 0.0446     \\
 $v_{\rm max}$$^{b}$ (\kms)& 16,821         & 16,109      & 13,722\\
 $v_{\rm min}$$^{c}$ (\kms)&  5285          & 4095        &  8131\\
 $AI$ $^{d}$  (\kms)        & 2726.1$\pm$2.6 & 1018.0$\pm$1.1  & - \\
 Observed depth$^{f}$ (\%)  & 33             & 19               &   5 \\
 Model depth$^{f}$ (\%)     & 33.9           & 18.1             &   7.6
\enddata
\tablenotetext{a}{Oscillator strengths are from NIST Atomic Spectra Database 
(http://physics.nist.gov/PhysRefData/ASD/).}
\tablenotetext{b}{Maximum blueshifted velocity of the BAL troughs from the emission line.}
\tablenotetext{c}{Minimum blueshifted velocity of the BAL troughs from the emission line. }
\tablenotetext{d}{The absorption index redefined by Zhang et al. (2010)}
\tablenotetext{e}{The absorption equivalent width of the BAL troughs.}
\tablenotetext{f}{Maximum depth of BAL troughs. Uncertainties on the observed depths are $\sim 5\%$.
The model depths are given for $\tau_{\rm H\alpha}=5.51\pm0.69$ and $C_f=34\%\mp2.0\%$.}
\end{deluxetable}

\figurenum{3}
 
\subsection{Further Confirmation of Balmer BALs}
As one know, \ion{Fe}{2} emission is  unique in individual quasars and  not exactly the same as 
the \mbox{I Zw 1} template. Furthermore, the ratio of \ion{Fe}{2} multiplets with different excitation
is very sensitive to  the temperature and density of the gas. 
The Balmer BAL troughs are unfortunately falling into the regimes of \ion{Fe}{2} multiplets.
Thus, we used a new \ion{Fe}{2} template derived from the Keck spectrum of IRAS 07598+650 (V{\'e}ron-Cetty et al. 2006)
to investigate the impact of \ion{Fe}{2} templates.
We first decompose the continuum emission of IRAS 07598+650 from the Keck spectrum 
trough a single power-law fitting based on the above-mentioned continuum windows in \S2.
After subtracting the modeled continuum, the residual spectrum which includes
broad and narrow \ion{Fe}{2} and Balmer emission lines is used as emission template. 
In panel (a) and (b) of Figure \ref{f3}, the cyan curves show the new best-fit model, 
which is the sum of power-law continuum (pink) and scaled emission template. 
It can be seen that the new unabsorbed model around H$\alpha$ and H$\beta$ BAL troughs is below the best-fitting
\mbox{I Zw 1} template causeing weaker H$\alpha$ and H$\beta$ absorption. 
Conversely, it shows a stronger H$\gamma$ BAL than the previous measurement.

In order to Further rule out the possibility that Balmer BALs are 
resulting from the artifacts of spectral fittings, we constructed a sample of 500 quasars whose spectra 
can match the slop and primary \ion{Fe}{2} multiplets, i.e., \ion{Fe}{2}$\lambda\lambda$ 4472-4731 \AA~
and \ion{Fe}{2}$\lambda\lambda$ 5169-5325 \AA~ of \J1523.
We used them to normalize the observed spectra of \J1523 
and calculated the absorption equivalent width of Balmer BALs. 
The average values are $2769.7\pm252.6$ \kms, $1341.4\pm217.1$ \kms
and $138.9\pm39.6$ \kms for H$\alpha$, H$\beta$ and H$\gamma$ BALs respectively. 
From the average values and dispersion of the absorption equivalent widths,
it can be seen that the uncertainty of \ion{Fe}{2} templates 
and/or the continuum determination can affect the measurement of BAL parameters, 
but the existence of Balmer BALs is unquestionable. 

\subsection{Optical Depth and Covering Factor}
Percentage absorption depths of the troughs (Table \ref{tab2}) decrease 
as the upper term of the transition increases. 
However, the decline in depth is less than the decrease in transition oscillator strengths 
($\sim 5$ from H$\alpha$ to H$\beta$), leading to the absorption saturated. %
There is a residual intensity of $\sim 2/3$ of the modelled flux at the velocity
of the maximum absorption depth (Figure \ref{f3}, panel (c)), suggesting that the absorption materials only obscure  
$\sim 1/3$ of the total continuum region. Theoretically,
the absorption  depth is defined as 
\begin{eqnarray}
D(v)=1-I(v)
\label{dep}
\end{eqnarray}
for a partially obscured absorber, where 
\begin{eqnarray}
I(v)=1-C_f(v)+C_f(v)e^{-\tau(v)}
\label{norf}
\end{eqnarray}
is the normalized intensity in the troughs, 
$C_f(v)$ is percentage covering factor of the absorber 
and $\tau(v)$ is the optical depth for the relevant ion (e.g., Hall et al. 2003).  
We can calculate $D_i(v)$ for each Balmer trough from the normalized spectrum, 
and the relative values of $\tau_i(v)$ are determined by the known oscillator strengths
\footnote{$\tau_2=R*\tau_1$, where $R=g_2f_2\lambda_2/g_1f_1\lambda_1$ is the ratio of the optical depth of
H$\beta$ to that of H$\alpha$, the $g_i$ are the statistical weights, the $f_i$ are the oscillator strengths, 
and the $\lambda_i$ are the wavelengths of the lines.}. 
Thus, we can estimate $C_f(v)$ and $\tau(v)$ through the observed H$\alpha$ and H$\beta$ troughs.
The maximum absorption depth of the Balmer lines is found to be 
$\tau_{\rm H\alpha}=5.51\pm0.69$ with coveing factor $C_f=34\%\mp2.0\%$ at the velocity of maximum depth (Table \ref{tab2}).

In panel (d) of Figure \ref{f3}, the covering factor $C_f(v)$ and the optical depth of 
H$\alpha$ absorption $\tau(v)$ are shown as a function of the blueshiftted velocity.
The absorption materials have the maximum covering factor at the velocity of the maximum absorption depth ($v\sim10,700$ \kms)
and smaller covering factors at the higher or lower velocities. 
The optical depths, ie., the column densities,
decrease as a function of the blueshifted velocity from $\tau\approx8.0$ at $v=5000$ \kms and $\tau\approx3.0$ at $v=16,000$ \kms. 
In panel(c) of Figure \ref{f3}, we compared the theoretical absorption troughs (blue curves) with
the observed normalized spectra for Balmer BALs. 
We find that the observed absorption trough for H$\gamma$ is shallower than the theoretical profile. 
The absorption of H$\gamma$ is relatively weak and difficult to accurately measure.

Using the optical depths derived above, we calculate the column densities of H I as a function of velocity 
using the general expression (e.g., Arav et al. 2001)
\begin{eqnarray}
N(\Delta v)= \frac{m_e c}{\pi e^2} \frac{1}{\lambda_0 f_{0}} \tau(\Delta v) 
= \frac{3.7679\times 10^{14}}{\lambda_0 f_{0}} \tau(\Delta v) \ \  {\rm cm^{-2}},
\label{colden}
\end{eqnarray}
where $\lambda_0=6564.41$ and $f_{0}=0.6400$ are the wavelength and the  oscillator strength of H$\alpha$, respectively.
The results are shown in panel (d) of Figure \ref{f3}. 
The total column density is obtained $N_{\rm H\ I}=5.52\pm0.27\times10^{15}$ cm$^{-2}$ by integrating equation \ref{colden}.

\subsection{Absorption of \ion{He}{1}*$\lambda$10830}

\figurenum{4}
\begin{figure}[tbp]
\epsscale{0.8} \plotone{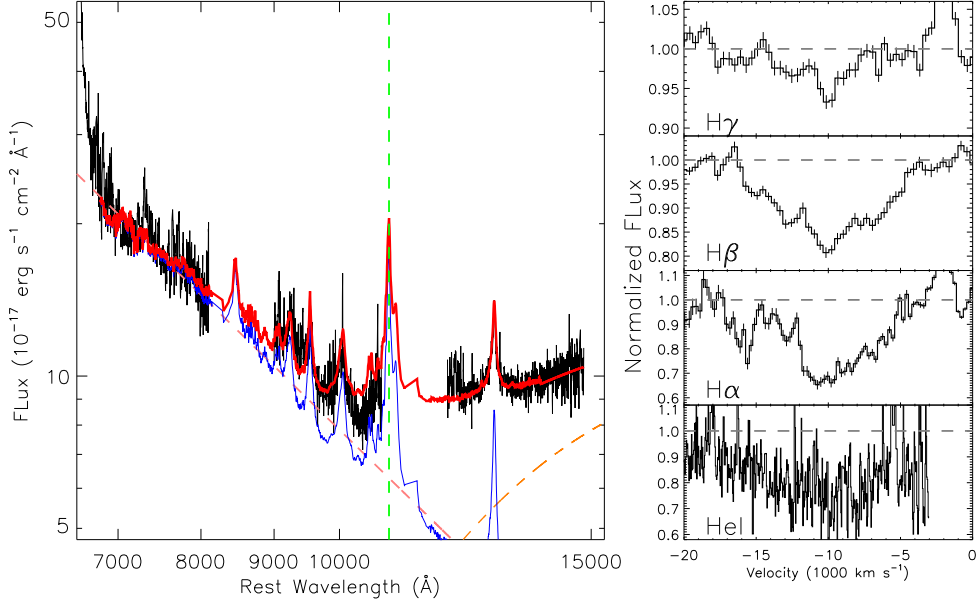}
\caption{Left: Observed spectra (black) of \ion{He}{1}*$\lambda$10830 regime overplayed 
with the best-fit model (red) for SDSS J1523+3914. 
Right: Normalized spectrum as a function of velocity of \ion{He}{1}$^* \lambda10830$. 
The normalized spectra of three Balmer BALs are shown for comparison.}
\label{f4}
\end{figure}

Only the blue wing of the redshifted \ion{He}{1}*$\lambda$10830 emission line 
is detected with the TripleSpec spectrum at H-band (Panel (a) of Figure \ref{f2}),
To obtain the unabsorbed spectral intensity around \ion{He}{1}*$\lambda$10830, we used the following procedure.
We downloaded the high-quality near-infrared broad emission line spectra presented by Landt et al. (2008)
and found four objects which can be compared with  \J1523: 3C 273, IRAS 1750, HE 1228+013 and PDS 456.
These spectra are similar to that of \J1523~in terms of emission lines, but different in their continuum slopes.  
Thus we could use line-free spectral regions in each comparison object to determine the continuum and thus the normalized emission line profiles. 
We focus on a limited wavelength range between 0.7 and 1.5 $\mu$m and fit the line-free continuum 
bands of each object with a single power-law for the accretion disk and a black-body emission component 
for hot dust emission. Then the emission line templates are obtained by dividing the observed spectra 
by the best-fitting continuum. Finally, we multiplied the emission line templates by a single power-law and
black-body emission to reconstruct the NIR spectrum of \J1523. 
IRAC 1750 provided the best-matched NIR emission line template with the minimized $\chi^2$.
In Figure \ref{f4}, left panel shows the observed spectra of \ion{He}{1}*$\lambda$10830 regime overplayed with 
the best-fit model for \J1523. The pink and orange lines represents the power-low and black-body emission components, respectively. 
For comparison, we overplotted the emission line template multiplied by a power-law component for IRAC 1750 (blue curve). 
Right-bottom panel shows the normalized spectrum of \ion{He}{1}* BAL trough 
obtained by dividing the observed spectrum by the best-fitting emission line plus continuum template. 
Trough the comparisons of \ion{He}{1}* and Balmer BALs, the velocity structure of \ion{He}{1}* BAL (black curve) is similar to those of Balmer BALs.

\section{Discussion}

\subsection{Comparison with Other Balmer absorption}
Table \ref{tab3} gives outflow velocities of the absorber for various Balmer BALs, as reported in literature. 
In three cases (SDSS J125942.80+121312.6, LBQS 1206+1052 and SDSS J222024.59+010931.2),
the absorption widths of Balmer absorption troughs are $\Delta v \gtrsim 1500$ \kms, and can be classified as Balmer BALs.
However, the widths for other quasars are only several hundred \kms, and could be classified as Balmer NALs. 
As shown in Table \ref{tab2}, the widths of Balmer BALs in \J1523\ 
are $\Delta v \simeq 12,000$ \kms, and are the broadest Balmer absorption lines which have ever been found. 
The redshift of the Balmer absorption troughs in \J1523 is $z_{absor}$=0.6039$\pm$0.0021, 
and the blueshifted velocity is $v=10,353$ \kms, which can even reach $v_{\rm max} \sim 17,000$ \kms\ with regard to the Balmer emission lines.  
It is a factor of two higher than  
the maximum blueshifted velocity from the previously known Balmer absorption lines. 
In fact, the blueshifted velocities in two-thirds of the known Balmer BALs are very small (only $\leq 1000$ \kms).
Compared with the above two Balmer BAL quasars with strong [\ion{O}{3}] emission in literature,
we find that the [\ion{O}{3}] line in \J1523\ is relatively weak (\mbox{$EW_{\rm [O~III]}= 3.46\pm0.12$ \AA}). 
\J1523 also has the weakest [O III] emission among nine AGNs with Balmer absorption lines.  
\J1523 and SDSS J222024.59+010931.2 
are obviously inconsistent with the previous assertion 
that Balmer BALs are found in FeLoBAL quasars with relatively strong [\ion{O}{3}] emission (Aoki et al. 2006; Hall et al. 2008).

Among the ten objects in literature and this work, 
absorption lines are usually detected in \ion{He}{1}* and \ion{Fe}{2}*.   
\ion{He}{1}* absorption lines arise from the metastable triplet level \ion{He}{1}* $2^3S$, 
which is populated by recombination from He$^+$ with electrons. 
Transitions from this level will generate a series of absorption lines at $\lambda\lambda$3189, 3889, 10830 \AA,
which are detected in seven quasars, except for SDSS J112611.63+425246.4, SDSS J125942.80+121312.6 and SDSS J172341.10+555340.5.
Another interesting fact is that seven out of ten Balmer absorption AGNs 
show abundant absorption lines arising from the excited \ion{Fe}{2} 
The co-occurrence may indicate that it is probable that these three absorption phenomena are closely related. 
Combined diagnostics of them can determine the density and the ionization state of the absorption gas, 
and put constraints on the geometry and physical conditions of outflows.
For example, the absorbers in SDSS J112526.12+002901.3 are considered to have the parameters of
${\rm log_{10}}~n_e~({\rm cm}^{-3})=9.25^{+0.5}_{-0.25}$, ${\rm log_{10}}~N_{\rm H}~({\rm cm}^{-2})=22^{+0.5}_{-0.25}$ 
and ${\rm log_{10}}~U= -1.75\pm0.25$. The derived distance from the central engine is about $1.8^{+1.4}_{-1.0}$ pc, 
which is about 10 times the size of broad emission line region and similar to the radius of the inner edge of dusty torus (Shi et al., in pre.).
These estimations provide us some enlightenment about the physical properties of the absorbers in \J1523.
\J1523 has the strongest Balmer BAL troughs discovered to date, and 
exhibits simultaneously the broad absorption lines of \ion{He}{1}* $\lambda$10830 and both ground and excited state \ion{Fe}{2}.
However, these absorption lines have extremely broad velocity structures and heavily overlap each other, 
which can not be trivially separated.

\subsection{Physical Properties of the Absorber}
\figurenum{5}
\begin{figure*}[tbp]
\epsscale{1.0} \plotone{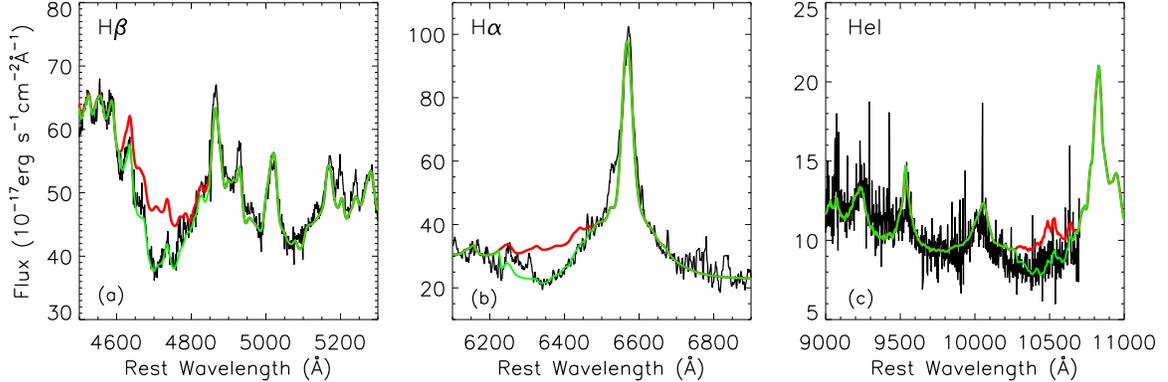}
\caption{Best-fitted results of the optimal photoionization models. 
Observed spectra (black curves) of H$\beta$ and H$\alpha$ and \ion{He}{1}*$\lambda$10830 regimes for \J1523\
overplayed with the best-fit  emission models (red  curves, also shown by red  curves in panel (a-b) of Figure \ref{f3} 
and panel (a) of Figure \ref{f4}). 
The green lines represent the model profiles of absorption lines.}\label{f5}
\end{figure*}

Since the absorption lines are blended, it is not trivial to obtain the geometry and physical conditions of outflow winds
via the combination of the absorption line diagnostics. 
The large-scale synthesis code CLOUDY (c10.00; Ferland et al. 1998) is employed  
to evaluate the absorption lines of these atoms/ions in the extensive parameter space. 
The geometry is assumed as a slab-shaped absorbing medium exposed to the ionizing continuum from the central engine 
with uniform density, metallicity and abundance pattern.
The full 371 levels Fe$^+$ model is used to reproduce the \ion{Fe}{2} absorption, 
and solar elemental abundance is adopted and the gas is assumed free of dust.
In addition, a typical AGN multi-component continuum is setted as incident ionizing radiation,
where the ``Big Bump" component peaks at $\approx$ 1 Ryd and is parameterized by T=$1.5\times10^5$ K.
The X-ray to UV ratio is $\alpha_{\rm ox}\sim-1.4$, and the low-energy slope of the Big Bump continuum is
$\alpha_{\rm UV}=-0.6818$ which is measured in \S 3.1. The slope of the X-ray component is setted to 
the default $\alpha_{\rm x}=-1$ (see details in Hazy, a brief introduction to CLOUDY C10; http://www.nublado.org).
We calculated a series of photoionization models with different ionization parameters, electron densities and 
hydrogen column densities. The ranges of parameters are $-3\leqslant{\rm log_{10}}~U\leqslant0$, 
$5\leqslant{\rm log_{10}}~n_{\rm e}~({\rm cm^{-3}})\leqslant11$ and 
$21\leqslant{\rm log_{10}}~N_{\rm H}~({\rm cm^{-2}})\leqslant24$ with a step of 0.5 dex.

 Synthetic model spectra are constructed to compare the simulations with observation. 
The underlying assumption is that other absorptions such as \ion{He}{1}$^*$ and \ion{Fe}{2} have the same profile as Balmer lines, 
which means for any ion the fraction of column density at given radial velocity to the integrated ionic column density is the same, 
and the covering factor as function of radial velocity is the same. Therefore in constructing the model spectra, 
for a given absorption line, the ionic column density predicted by CLOUDY on the lower level of the transition is distributed to 
different outflow velocities following the fractional column density distribution versus $v$ from Balmer series, 
evaluating the optical depth $\tau$ as function of $v$.  And then considering the effect of partial covering as 
$f_{\rm model}=  C_f*e^{-\tau} f_0 + (1-C_f) f_0$ to get the model absorption profile, 
where $f_0$ is the template for unabsorbed background radiation field.  
When comparing the model spectrum ($f_{\rm model}$) with the observed one,
the spectrum from 4000 to 7000 \AA~ covering the three Balmer BALs are employed in the fitting process.
$f_0$ is actually the unabsorbed intrinsic spectrum, for example, the red curves shown in top panels of 
Figure \ref{f3} for Balmer lines. For each set of parameters ($U$, $n_{\rm e}$ and $N_{\rm H}$),
we calculated all possible combinations of individual models, and selected the one with minimized $\chi^2$  
which is plotted in panel (a) and (b) of Figure \ref{f5} (green curves). We derived physical parameters 
${\rm log_{10}}~U=-1.0$, ${\rm log_{10}}~n_{\rm e}~({\rm cm^{-3}})=9$ and
${\rm log_{10}}~N_{\rm H}~({\rm cm^{-2}})=23.5$ for Balmer BALs.
The green curves represent the model profiles of H$\beta$ and H$\alpha$ absorption 
which match well to the observed spectra. Meanwhile, we also plotted the model absorption trough
of \ion{He}{1}$^*\lambda10830$ in panel (c) of Figure \ref{f5}.  
The model absorption trough of \ion{He}{1}$^*\lambda10830$ BAL is consistent with observation.
Note that, $f_0$ in the \ion{He}{1}$^*\lambda$10830 region is the blue curve shown in panel (a) of Figure \ref{f4},
and the modeled $f_{\rm model}$ should be added with the black-body emission component 
(orange dashed line in Figure \ref{f4}) in the comparison of the model spectrum  with observation.

\begin{figure}[tbp]
\epsscale{0.5} \plotone{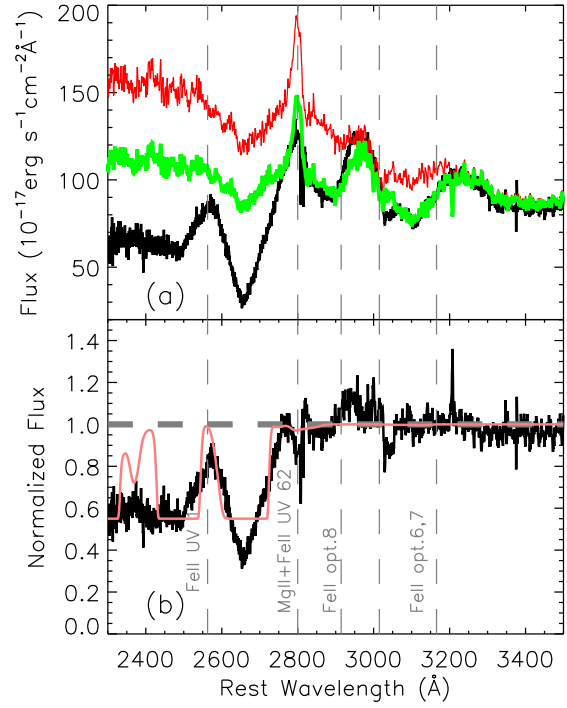}
\caption{Panel (a): Observed spectra (black curve) of \ion{Mg}{2} and \ion{Fe}{2} absorption regimes overplayed with the comparison spectrum of SDSS J082806.18+063608.3 (red curve). The green lines represent the model profiles of absorption lines. Panel (b): Residual absorption components of \ion{Mg}{2} and \ion{Fe}{2} absorption. The brown lines means the broadened  and blueshifted low density model.
}\label{f6}
\end{figure}

We further explored the various \ion{Fe}{2} absorption and \ion{Mg}{2} doublets in the SDSS spectrum.
Here, the comparison spectrum of SDSS J082806.18+063608.3 was used to represent the unabsorbed intrinsic spectrum. 
The  observed spectrum (black curve), and modeled spectrum with absorption (green curve) for 
\J1523~are plotted in panel (a) of Figure \ref{f6}.
It can be seen that the modeled spectrum with absorption  is in good agreement with the observed one at the red wing of \ion{Mg}{2}
and the longer wavelengths. The normalized fluxes at wavelength longger than 2800 \AA~ are around 1.
That means that the absorption multiplets of \ion{Fe}{2} opt.6,7 and opt.8 observed in \J1523 can be largely recovered
using the above modeled absorption.  
However, the match of \ion{Fe}{2} opt.6,7 and opt.8 between model and observation is still fluky.
\J1523~is observed to decrease in flux on the timescale of 8.99 years 
from the FBQS to the SDSS, but the absorption of \ion{Fe}{2} opt.6,7 and opt.8 
changes weakly, and the dramatic absorption variabilities are \ion{Fe}{2} UV 1, UV 62 and \ion{Mg}{2}.
A photoionization model with lower density (${\rm log_{10}}~n_{\rm e}~({\rm cm^{-3}})=7$) 
and ionization parameter ( ${\rm log_{10}}~U=-2.0$) than the above can approximatively
match the absorption variations (see Figure 2 of Zhang et al. 2015b). 

For the SDSS blue-side spectrum, there seems to be other absorption components for \ion{Fe}{2} UV 1, UV 62 
and \ion{Mg}{2}, corresponding to outflow materials with low density. Indeed, the comparison
of \J1523 and the two templates in panel (b) of Figure \ref{f2} suggests the existence of the other absorption components.
\ion{Fe}{2} opt.6,7 and opt.8 have similar absorption depth as the Balmer absorption troughs,
while the troughs of \ion{Fe}{2} UV 1 are nearly twice as deep as \ion{Fe}{2} opt.6,7 
and opt.8 multiplets. Moreover, the residual fluxes of \ion{Fe}{2} UV 62 and \ion{Mg}{2} BAL troughs are even only one-third of 
\ion{Fe}{2} opt.6,7 and opt.8 troughs. We broadened a low density model 
(${\rm log_{10}}~U=-1.0$, ${\rm log_{10}}~n_{\rm e}~({\rm cm^{-3}})=5$ and ${\rm log_{10}}~N_{\rm H}~({\rm cm^{-2}})=23.5$)
with a single Gaussian profile and blueshifted it to approximatively model the residual component shown in panel (b) of Figure \ref{f6}.
The FWHM of Gaussian profile is 3000 \kms and the blueshifted velocity is 14,000 \kms. 
If we use a uniform covering factor for this component,  $C_f$ is found to be $\sim45\%$. 
 We also noticed that the residual absorption spectrum indicate stronger absorption of  \ion{Fe}{2} UV144-149 and UV158-164 multiplets (around 2400 \AA)  than the optical model (Panel (b) of Figure \ref{f2}), which most likely suggests a higher density or some microturbulence in the outflow winds (Shi et al. in pre.).

\figurenum{6}

\subsection{Black Hole Mass Estimate and Location of the Absorber }
From the spectral fittings in Section 3.1, we can obtain the continuum and emission line parameters,
and derive the quasar fundamental parameters. 
The measured monochromatic luminosity at 5100 \AA~in the rest frame is $L_{5100}=9.29\times10^{44}$ erg s$^{-1}$.
The bolometric luminosity $L_{\rm bol}$ is estimated from the luminosity $L_{5100}$
with a bolometric correction of 9 (Kaspi et al. 2000).
The full width at half-maximum (FWHM) of broad Balmer emission lines is 1744 \kms, and then
the BH mass $M_{\rm BH}=7.8\times10^7$ M$_{\sun}$, is estimated using 
the following prescription based on the luminosity at 5100\AA~and the width of H$\beta$ emission 
(Vestergaard \& Peterson 2006):
\begin{equation}
\log \,M_{\rm BH} =
\log \,\left[ \left(\frac{\rm FWHM}{\rm 1000~km~s^{-1}}
\right)^2 ~
\left( \frac{L_{\rm 5100}}{10^{44}
\rm erg~s^{-1}}\right)^{0.50} \right] 
+ (6.91 \pm 0.02).
\label{Mbh}
\end{equation}
The corresponding Eddington ratio is $L_{\rm Edd}$, $l_{\rm E}=L_{\rm bol}/L_{\rm Edd}=0.85$.
The black hole mass computing formula of McLure \& Dunlop (2004) give the similar results, 
$M_{\rm BH}=5.76\times10^7$ M$_{\sun}$ and $l_{\rm E}=1.15$.
\J1523 is a typical narrow-line Seyfert 1 galaxy (NLS1) with a low black hole mass but high accretion ratio.

The radius of broad emission line regions (BELRs), $R_{\rm BLR}$,
can be estimated using the formula based on the luminosity at 5100\AA,
\begin{eqnarray}
R_{\rm BLR} = \alpha \left(\dfrac{L_{5100}}{10^{44}~\rm erg~s^{-1}}\right)^{\beta}~{\rm lt-days,}
\label{rblr}
\end{eqnarray}
where the parameters, $\alpha$ and $\beta$ are $30.2\pm1.4$ and $0.64\pm0.02$ given in Greene \& Ho (2005) 
and $20.0^{+2.8}_{-2.4}$ and $0.67\pm0.07$ given in Kaspi et al. (2005) respectively.
Thus, the luminosity yields $R_{\rm BLR}\approx0.1\pm0.02$ pc.
The radius of the dust torus, $R_{\rm Torus}$, can also be estimated based on the thermal equilibrium 
of the inner side of the torus as
\begin{eqnarray}
R_{\rm Torus}=\sqrt{\dfrac{L_{\rm bol}}{4\pi\sigma T^4}}, 
\label{torus}
\end{eqnarray}
where $\sigma$ is the Stefan-Boltzmann constant, $T (\sim 1500~\rm K)$ is the temperature of inner side of the tours.
Then, we get $R_{\rm Torus}\approx 0.5$ pc.

Ionization parameter is defined as:
\begin{eqnarray}
U = \int^{inf}_{\nu_0} \dfrac{L_{\nu}}{4\pi r^2 h\nu n_{\rm e}c}~d\nu=\dfrac{Q}{4\pi r^2 n_{\rm e}c},
\label{ioni}
\end{eqnarray}
in which,  $\nu_0$ is the frequency corresponding the hydrogen edge,
and $Q$ is the ionization photons emission rate.
Using above ionization equation and the inferred $Q$, $U$ and $n_{\rm e}$ values, 
one can get the distance of the absorption gas of Balmer BALs from central ionization source, $R_{BAL}\approx0.2$ pc.
Thus, the high density outflow gas of Balmer BALs locates at or outside of the BELRs, 
and less than the distance of the dust torus.
Similarly, the estimated distance of the low density outflow gas is much farther than the outflow materials with high density,
reacheing several tens of parsecs.

\subsection{Origin of the Outflow Winds}

\begin{deluxetable}{ccc cccc }
\tabletypesize{\scriptsize}
\rotate
\tablecaption{Absorption Parameter Comparison
\label{tab3} }
\tablewidth{0pt}
\tablehead{
\colhead{Target}&
\colhead{$V_{blueshift}^{\dagger}$ }&
\colhead{Absorption Width} &
\colhead{Inferred Density}  &
\colhead{Inferred Distance}  &
\colhead{References}&
\colhead{Notes}\\
\colhead{}&
\colhead{(\kms)}&
\colhead{(\kms)}&
\colhead{(cm$^{-3}$)} &
\colhead{(pc)} &
\colhead{}&
\colhead{}
}
\startdata
NGC 4151                            &$\sim 1000$& $\sim$350 &             &         & Hutchings et al. (2002) & \\
SDSS J112526.12+002901.3$^{\ddagger}$&$71.9\pm7.2^b$, $-651.2\pm13.7^r$&$199.4\pm16.4^b$, $398.6\pm32.6^r$&10$^{9.25^{+0.5}_{-0.25}}$& 1.8$^{+1.4}_{-1.0}$&Hall et al. (2002)&mntFeLoBAL\\
SDSS J083942.11+380526.3$^{\ddagger}$& 520      & $\sim$340 & 10$^{8.25}$ & 4-40    & Aoki et al. (2006) &  \\
SDSS J125942.80+121312.6$^{\ddagger}$& 3400     &$2000\pm200$& 10$^9$     & 1       & Hall (2007)& BAL\\
SDSS J102839.11+450009.4            & 670       &  149$\pm$7&             &         & Wang et al. (2008) & \\
SDSS J172341.10+555340.5            & 5370      &450$\pm$130&             &         & Aoki et al. (2010) & mntFeLoBAL\\
LBQS 1206+1052						& 726       & 2000      & 10$^{6-8}$  &         & Ji et al. (2012)   & BAL \\
SDSS J222024.59+010931.2			& 0         & $\sim$1500& 10$^6$	  &         & Ji et al. (2013)   & BAL \\
SDSS J112611.63+425246.4 			& $300\pm70$&$290\pm 40$&             &         & Wang \& Xu (2015)  &  \\
SDSS J152350.42+391405.2			&10,353     &$\sim$12,000&10$^9$ 	  & 0.2     & this paper & BAL
\enddata
\tablenotetext{\dagger}{Negative value means redshift.}
\tablenotetext{\ddagger}{The parameter estimatian comes from Shi et al. (in pre.).}
\tablenotetext{b, r}{The blue and red component of the absorption profile. }
\end{deluxetable}

In \S 4.1, we compared Balmer BALs in \J1523 with non-stellar Balmer absorption lines in previous literature,
and found that \J1523 presents the broadest Balmer absorption lines with the maximum blueshifted velocity. 
Then, a question is naturally raised: why does this object present such unique absorption property?
In Zhang et al. (2014), we know that outflow velocity strongly or moderately depends on the Eddington ratio, luminosity,
UV and NIR slopes (also see Hamann 1998; Misawa et al. 2007; Ganguly et al. 2007).
The first three items represent the ionization SED and the amount of the high-energy photons,
and the last item is the possible contribution of the dust to the outflow acceleration.
\J1523 is a high-luminous NLS1 with near-Eddington accretion rate.
Meanwhile, the NIR slope of \J1523 is $\beta_{\rm NIR}\sim 0.8$, which is redder than those of most BAL quasars (see Figure 2 in Zhang et al. 2014).
These statistical properties bring about the occurrence of the high-velocity outflow.

From Table \ref{tab3}, we knew that only three of literature quasars are the Balmer BALs, and others are just the Balmer NALs.
Indeed, the absorption troughs of Balmer and \ion{He}{1}$^*$ series can help us to straighten out the classification of BALs.
For example, there are suspected overlapping absorption features of \ion{Fe}{2} in two objects, 
Hall et al. (2002) classified them into a special absorption subtype, the so-called many-narrow-trough FeLoBAL (mntFeLoBAL).
However, the troughs of Balmer and \ion{He}{1}$^*$ series present the true absorption profiles with narrow widths,
and these profiles are used to model \ion{Fe}{2} absorption multiplets (Shi et al. in pre.). 
The CLOUDY simulations provided the physical conditions of outflow gases for part of objects listed in Table \ref{tab3}.
It can be seen that almost all absorption lines are constrained to come from 
high density gases ($\sim 10^6 - 10^9\ {\rm cm}^{-3}$),
and \J1523 is among sources with the highest density except SDSS J112526.12+002901.3\footnote{For SDSS J112526.12+002901.3,
the derived distance from the central engine is about 10 times the size of broad emission line region and 
similar to the radius of the inner edge of dusty torus. 
 Shi et al. suggest that the failed disk wind cloud and infalling torus clump are more likely the origin of 
such blueshifted and redshifted absorbing gas.}.
The high density suggests that the outflow winds should survive in the inner region of the AGN.
The photoionization model gave an estimation of  a distance of $\sim 0.2$ pc. 
The outflow winds in \J1523 locate at such close from the central ionizing source which is slightly farther than that of BELRs.
In the disk wind scenarios (Murray et al. 1995), most photoionized clouds emerge from the accretion disk and accelerate outwards.
The velocities of the clouds are assumed to be a function of the initial velocities, the terminal velocities 
and the radii that they are away from the center. The terminal velocities are approximately inversely proportional 
to the square root of the radius at which the streamlines. 
 Thus the innermost streamlines have the highest rotational and terminal radial velocities and highest ionization states. 
Balmer absorption winds in \J1523\ are just located at the distance of $\sim 0.2$ pc from the central ionizing source.
We suggest that is the reason why Balmer BALs in this objetc have very large outflow velocities. 
Panel (c) of Figure \ref{f2} shows that the minimum velocity of the BAL troughs is $v_{\rm min}\sim 5000$ \kms, 
that means the absorption winds have large initial velocities when they emerge, or they have been accelerated from (inside) the BELRs.
If the mass flux is continuous, the optical depths decrease as a function of the blueshifted velocity. 
If the outflow is gathering speed, it seems that the volumes of absorption winds expand gradually during acceleration outwards.
That is consistent with the existence of the excess outflow component of 
\ion{Fe}{2} UV 62 and \ion{Mg}{2} BAL troughs (Figure \ref{f6}).
This component shows higher blueshifted velocity ($v\sim 14,000$ \kms), 
larger covering factor ($C_f\sim45\%$) and lower density (${\rm log_{10}}\ n_e\ {\rm (cm^{-3})}=5$), 
and is estimated to exist in remoter regions (several tens of parsecs).

\section{Conclusion}

In this work, we present the discovery of Balmer-series absorption lines from H$\alpha$, H$\beta$ and H$\gamma$ in \J1523 
from the quasi-simultaneous optical and near-infrared spectroscopy. 
The redshift of the Balmer absorption troughs is $z_{\rm absor}=0.6039\pm0.0021$, 
and it is blueshifted by $v = 10,353$ \kms with regard to the Balmer emission lines. 
Balmer BALs have outflowing velocities a factor of two than the previously known Balmer absorption lines.
We searched for the same velocity components seen in other NIR Balmer absorption lines. 
We found a component in the \ion{He}{1}$^*\lambda10830$ absorption line at the same redshift.
The absorption trough in \ion{He}{1}$^*\lambda10830$ has a uniform absorption profile with the Balmer-series, 
with the absorption width $\Delta v \sim 12,000$ \kms. 
Therefore \J1523 is the object with the broadest Balmer absorption lines detected so far.
We measured the profiles of Balmer BELs and derived their widths of $FWHM\sim 1744$ \kms.
The estimation of the fundamental parameters shows that \J1523 is a typical 
NLS1 with a low black hole mass but high accretion ratio.
We approximately evaluate the BALs by employing the CLOUDY in a extensive parameter space.
The outflow winds of Balmer BALs are suggested to be of the electron density of ${\rm log_{10}}\ n_e\ {\rm (cm^{-3})}=9$, 
an ionization parameter of ${\rm log_{10}}\ U=-1$, 
and the distance of $\sim 0.2$ pc from the central ionizing source which is slightly farther than that of BELRs.

\acknowledgments
Many thanks to  the anonymous referee for the helpful suggestions.  Many thanks to Dr. Huiyuan Wang and Dr. Xueguang Zhang for very helpful discussions. This work is supported by Chinese Natural Science Foundation (NSFC-11203021, 11573024), National Basic Research Program of China (``973'' Program, 2013CB834905) the SOC Program (CHINARE-2015-02-03). 

We acknowledge the use of the Hale 200-inch Telescope at Palomar Observatory through the Telescope Access Program (TAP) and the Antarctic bright star survey telescope (BSST), as well as the archive data from the FIRST, NVSS, SDSS, WISE and Catalina Surveys. The BSST is funded by the ``Antarctic Inland Scientific Research Station'' Program and the SOC Program (CHINARE-2015-02-03). TAP is funded by the Strategic Priority Research Program ``The Emergence of Cosmological Structures'' (XDB09000000), National Astronomical Observatories, Chinese Academy of Sciences, and the Special Fund for Astronomy from the Ministry of Finance. Observations obtained with the Hale Telescope at Palomar Observatory were obtained as part of an agreement between the National Astronomical Observatories, Chinese Academy of Sciences, and the California Institute of Technology. The Catalina Surveys consist of the Catalina Sky Survey (CSS) and the Catalina Real-time Transient Survey (CRTS). The CSS survey is funded by the National Aeronautics and Space Administration under Grant No. NNG05GF22G issued through the Science Mission Directorate Near-Earth Objects Observations Program.  The CRTS survey is supported by the U.S.~National Science Foundation under grants AST-0909182. Funding for SDSS-III has been provided by the Alfred P. Sloan Foundation, the Participating Institutions, the National Science Foundation, and the U.S. Department of Energy Office of Science. The SDSS-III web site is http://www.sdss3.org/.

\clearpage

\begin{thebibliography}{}
\bibitem[Adelman-McCarthy et al.(2007)]{2007ApJS..172..634A} Adelman-McCarthy, J.~K., Ag{\"u}eros, M.~A., Allam, S.~S., et al.\ 2007, \apjs, 172, 634 
\bibitem[Ai et al.(2010)]{2010ApJ...716L..31A} Ai, Y.~L., Yuan, W., Zhou, H.~Y., et al.\ 2010, \apjl, 716, L31 
\bibitem[Antonuccio-Delogu \& Silk(2010)]{2010ASPC..427..343A} Antonuccio-Delogu, V., \& Silk, J.\ 2010, Accretion and Ejection in AGN: a Global View, 427, 343 
\bibitem[Aoki et al.(2006)]{2006ApJ...651...84A} Aoki, K., Iwata, I., Ohta, K., et al.\ 2006, \apj, 651, 84 
\bibitem[Aoki(2010)]{2010PASJ...62.1333A} Aoki, K.\ 2010, \pasj, 62, 1333 
\bibitem[Arav et al.(2001)]{2001ApJ...561..118A} Arav, N., de Kool, M., Korista, K.~T., et al.\ 2001, \apj, 561, 118 
\bibitem[Baskin et al.(2013)]{2013MNRAS.432.1525B} Baskin, A., Laor, A., \& Hamann, F.\ 2013, \mnras, 432, 1525 
\bibitem[Becker et al.(1995)]{1995ApJ...450..559B} Becker, R.~H., White, R.~L., \& Helfand, D.~J.\ 1995, \apj, 450, 559
\bibitem[Becker et al.(2000)]{2000ApJ...538...72B} Becker, R.~H., White, R.~L., Gregg, M.~D., et al.\ 2000, \apj, 538, 72
\bibitem[Brandt et al.(2000)]{2000ApJ...528..637B} Brandt, W.~N., Laor, A., \& Wills, B.~J.\ 2000, \apj, 528, 637 
\bibitem[Brinkmann et al.(1999)]{1999A&A...345...43B} Brinkmann, W., Wang, T., Matsuoka, M., \& Yuan, W.\ 1999, \aap, 345, 43
\bibitem[Condon et al.(1998)]{1998AJ....115.1693C} Condon, J.~J., Cotton, W.~D., Greisen, E.~W., et al.\ 1998, \aj, 115, 1693
\bibitem[Dong et al.(2008)]{2008MNRAS.383..581D} Dong, X., Wang, T., Wang, J., et al.\ 2008, \mnras, 383, 581 
\bibitem[Drake et al.(2014)]{2014ApJS..213....9D} Drake, A.~J., et al.\ 2014, \apjs, 213, 9
\bibitem[Fan et al.(2009)]{2009ApJ...690.1006F} Fan, L.~L., Wang, H.~Y., Wang, T., et al.\ 2009, \apj, 690, 1006 
\bibitem[Ferland et al.(1998)]{1998PASP..110..761F} Ferland, G.~J., Korista, K.~T., Verner, D.~A., et al.\ 1998, \pasp, 110, 761 
\bibitem[Forster et al.(2001)]{2001ApJS..134...35F} Forster, K., Green, P.~J., Aldcroft, T.~L., et al.\ 2001, \apjs, 134, 35 
\bibitem[Gallagher et al.(2002)]{2002ApJ...567...37G} Gallagher, S.~C., Brandt, W.~N., Chartas, G., \& Garmire, G.~P.\ 2002, \apj, 567, 37 
\bibitem[Gallagher et al.(2006)]{2006ApJ...644..709G} Gallagher, S.~C., Brandt, W.~N., Chartas, G., et al.\ 2006, \apj, 644, 709 
\bibitem[Ganguly et al.(2007)]{2007ApJ...665..990G} Ganguly, R., Brotherton, M.~S., Cales, S., et al.\ 2007, \apj, 665, 990 
\bibitem[Gibson et al.(2008)]{2008ApJ...675..985G} Gibson, R.~R., Brandt, W.~N., Schneider, D.~P., \& Gallagher, S.~C.\ 2008, \apj, 675, 985 
\bibitem[Gibson et al.(2009)]{2009ApJ...692..758G} Gibson, R.~R., Jiang, L., Brandt, W.~N., et al.\ 2009, \apj, 692, 758 
\bibitem[Gibson et al.(2010)]{2010ApJ...713..220G} Gibson, R.~R., Brandt, W.~N., Gallagher, S.~C.,  et al.\ 2010, \apj, 713, 220 
\bibitem[Granato et al.(2004)]{2004ApJ...600..580G} Granato, G.~L., De Zotti, G., Silva, L., Bressan, A., \& Danese, L.\ 2004, \apj, 600, 580
\bibitem[Green et al.(1995)]{1995ApJ...450...51G} Green, P.~J., Schartel, N., Anderson, S.~F., et al.\ 1995, \apj, 450, 51 
\bibitem[Greene \& Ho(2005)]{2005ApJ...630..122G} Greene, J.~E., \& Ho, L.~C.\ 2005, \apj, 630, 122 
\bibitem[Hall(2007)]{2007AJ....133.1271H} Hall, P.~B.\ 2007, \aj, 133, 1271 
\bibitem[Hall et al.(2002)]{2002ApJS..141..267H} Hall, P.~B., Anderson, S.~F., Strauss, M.~A., et al.\ 2002, \apjs, 141, 267
\bibitem[Hall et al.(2003)]{2003ApJ...593..189H} Hall, P.~B., Hutsem{\'e}kers, D., Anderson, S.~F., et al.\ 2003, \apj, 593, 189 
\bibitem[Hall et al.(2011)]{2011MNRAS.411.2653H} Hall, P.~B., Anosov, K., White, R.~L., et al.\ 2011, \mnras, 411, 2653 
\bibitem[Hamann(1998)]{1998ApJ...500..798H} Hamann, F.\ 1998, \apj, 500, 798
\bibitem[Hopkins et al.(2008)]{2008ApJS..175..356H} Hopkins, P.~F., Hernquist, L., Cox, T.~J., \& Kere{\v s}, D.\ 2008, \apjs, 175, 356 
\bibitem[Hutchings et al.(2002)]{2002AJ....124.2543H} Hutchings, J.~B., Crenshaw, D.~M., Kraemer, S.~B., et al.\ 2002, \aj, 124, 2543 
\bibitem[Ji et al.(2012)]{2012RAA....12..369J} Ji, T., Wang, T.-G., Zhou, H.-Y., \& Wang, H.-Y.\ 2012, Research in Astronomy and Astrophysics, 12, 369
\bibitem[Ji et al.(2013)]{2013ChA&A..37...17J} Ji, T., Zhou, H.-Y., Wang, T.-G., \& Wang, H.-Y.\ 2013, ChA\&A, 37, 17 
\bibitem[Kaspi et al.(2000)]{2000ApJ...533..631K} Kaspi, S., Smith, P.~S., Netzer, H., et al.\ 2000, \apj, 533, 631 
\bibitem[Kaspi et al.(2005)]{2005ApJ...629...61K} Kaspi, S., Maoz, D., Netzer, H., et al.\ 2005, \apj, 629, 61 
\bibitem[Kraemer et al.(2001)]{2001ApJ...551..671K} Kraemer, S.~B., Crenshaw, D.~M., Hutchings, J.~B., et al.\ 2001, \apj, 551, 671 
\bibitem[Krongold et al.(2010)]{2010ApJ...724L.203K} Krongold, Y., Binette, L., \& Hern{\'a}ndez-Ibarra, F.\ 2010, \apjl, 724, L203 
\bibitem[Landolt(2009)]{2009AJ....137.4186L} Landolt, A.~U.\ 2009, \aj, 137, 4186 
\bibitem[Landt et al.(2008)]{2008ApJS..174..282L} Landt, H., Bentz, M.~C., Ward, M.~J., et al.\ 2008, \apjs, 174, 282 
\bibitem[Laor \& Brandt(2002)]{2002ApJ...569..641L} Laor, A., \& Brandt, W.~N.\ 2002, \apj, 569, 641
\bibitem[McLure \& Dunlop(2004)]{2004MNRAS.352.1390M} McLure, R.~J., \& Dunlop, J.~S.\ 2004, \mnras, 352, 1390 
\bibitem[Liu et al.(2015)]{2015ApJS..217...11L} Liu, W.-J., Zhou, H., Ji, T., et al.\ 2015, \apjs, 217, 11 
\bibitem[Misawa et al.(2007)]{2007ApJS..171....1M} Misawa, T., Charlton, J.~C., Eracleous, M., et al.\ 2007, \apjs, 171, 1 
\bibitem[Murray et al.(1995)]{1995ApJ...451..498M} Murray, N., Chiang, J., Grossman, S.~A., \& Voit, G.~M.\ 1995, \apj, 451, 498 
\bibitem[Reichard et al.(2003)]{2003AJ....126.2594R} Reichard, T.~A., Richards, G.~T., Hall, P.~B., et al.\ 2003, \aj, 126, 2594 
\bibitem[Scannapieco \& Oh(2004)]{2004ApJ...608...62S} Scannapieco, E., \& Oh, S.~P.\ 2004, \apj, 608, 62 
\bibitem[Schlegel et al.(1998)]{1998ApJ...500..525S} Schlegel, D.~J., Finkbeiner, D.~P., \& Davis, M.\ 1998, \apj, 500, 525 
\bibitem[Skrutskie et al.(2006)]{2006AJ....131.1163S} Skrutskie, M.~F., Cutri, R.~M., Stiening, R., et al.\ 2006, \aj, 131, 1163 
\bibitem[Stoughton et al.(2002)]{2002AJ....123..485S} Stoughton, C., Lupton, R.~H., Bernardi, M., et al.\ 2002, \aj, 123, 485 
\bibitem[Trump et al.(2006)]{2006ApJS..165....1T} Trump, J.~R., Hall, P.~B., Reichard, T.~A., et al.\ 2006, \apjs, 165, 1 
\bibitem[Vacca et al.(2003)]{2003PASP..115..389V} Vacca, W.~D., Cushing, M.~C., \& Rayner, J.~T.\ 2003, \pasp, 115, 389 
\bibitem[V{\'e}ron-Cetty et al.(2004)]{2004A&A...417..515V} V{\'e}ron-Cetty, M.-P., Joly, M., \& V{\'e}ron, P.\ 2004, \aap, 417, 515
\bibitem[V{\'e}ron-Cetty et al.(2006)]{2006A&A...451..851V} V{\'e}ron-Cetty, M.-P., Joly, M., V{\'e}ron, P., et al.\ 2006, \aap, 451, 851 
\bibitem[Vestergaard \& Peterson(2006)]{2006ApJ...641..689V} Vestergaard, M., \& Peterson, B.~M.\ 2006, \apj, 641, 689 
\bibitem[Wang et al.(1999)]{1999ApJ...519L..35W} Wang, T.~G., Wang, J.~X., Brinkmann, W., \& Matsuoka, M.\ 1999, \apjl, 519, L35 
\bibitem[Wang et al.(2008)]{2008ApJ...674..668W} Wang, T., Dai, H., \& Zhou, H.\ 2008, \apj, 674, 668 
\bibitem[Wang et al.(2013)]{2013ApJ...776L..15W} Wang, H., Xing, F., Zhang, K., et al.\ 2013, \apjl, 776, L15 
\bibitem[Wang \& Xu(2015)]{2015A&A...573A..15W} Wang, J., \& Xu, D.~W.\ 2015, \aap, 573, AA15 
\bibitem[Weymann et al.(1991)]{1991ApJ...373...23W} Weymann, R.~J., Morris, S.~L., Foltz, C.~B., \& Hewett, P.~C.\ 1991, \apj, 373, 23 
\bibitem[Wilson et al.(2004)]{2004SPIE.5492.1295W} Wilson, J.~C., Henderson, C.~P., Herter, T.~L., et al.\ 2004, \procspie, 5492, 1295 
\bibitem[Wright et al.(2010)]{2010AJ....140.1868W} Wright, E.~L., Eisenhardt, P.~R.~M., Mainzer, A.~K., et al.\ 2010, \aj, 140, 1868 
\bibitem[Zhang et al.(2010)]{2010ApJ...714..367Z} Zhang, S., Wang, T.-G., Wang, H., et al.\ 2010, \apj, 714, 367 
\bibitem[Zhang et al.(2011)]{2011RAA....11.1163Z} Zhang, S.-H., Wang, H.-Y., Zhou, H.-Y., Wang, T.-G., \& Jiang, P.\ 2011, Research in Astronomy and Astrophysics, 11, 1163 
\bibitem[Zhang et al.(2014)]{2014ApJ...786...42Z} Zhang, S., Wang, H., Wang, T., et al.\ 2014, \apj, 786, 42
\bibitem[Zhang et al.(2015)]{2015ApJ...802...92Z} Zhang, S., Ge, J., Jiang, P., et al.\ 2015a, \apj, 802, 92
\bibitem[Zhang et al.(2015)]{2015ApJ...803...58Z} Zhang, S., Zhou, H., Wang, T., et al.\ 2015b, \apj, 803, 58 
\bibitem[Zhou et al.(2006)]{2006ApJ...639..716Z} Zhou, H., Wang, T., Wang, H., Wang, J., Yuan, W., \& Lu, Y.\ 2006, \apj, 639, 716
\end{thebibliography}
\end{document}